\documentclass{article}

\usepackage{graphicx}

\usepackage{microtype}
\usepackage{graphicx}
\usepackage{subfigure}
\usepackage{booktabs} % for professional tables
\usepackage{amsmath}
\usepackage{multirow} 

\usepackage{xurl}

\usepackage{hyperref}

\usepackage[accepted]{mlsys2025}

\mlsystitlerunning{Holistic Evaluation of State-of-the-Art LLMs for Code Generation}

\begin{document}

\twocolumn[
\mlsystitle{Holistic Evaluation of State-of-the-Art LLMs\\for Code Generation}

% unless you've provided the [accepted] option to the mlsys2025


\mlsyssetsymbol{equal}{*}

\begin{mlsysauthorlist}
\mlsysauthor{Le Zhang}{}
\mlsysauthor{Suresh Kothari}{}
\end{mlsysauthorlist}

% \mlsysaffiliation{to}{Department of Computation, University of Torontoland, Torontoland, Canada}

% \mlsyscorrespondingauthor{Cieua Vvvvv}{c.vvvvv@googol.com}

\mlsyskeywords{Machine Learning, Large Language Model, LLM, Code Generation, Empirical Evaluation, Algorithmic Efficiency, Prompt Engineering}

\vskip 0.05in

\begin{abstract}
This study presents a comprehensive empirical evaluation of six state-of-the-art large language models (LLMs) for code generation, including both general-purpose and code-specialized models. Using a dataset of 944 real-world LeetCode problems across five programming languages, we assess model performance using rigorous metrics: compile-time errors, runtime errors, functional failures, and algorithmic suboptimalities. The results reveal significant performance variations, with DeepSeek-R1 and GPT-4.1 consistently outperform others in terms of correctness, efficiency, and robustness. Through detailed case studies, we identify common failure scenarios such as syntax errors, logical flaws, and suboptimal algorithms, highlighting the critical role of prompt engineering and human oversight in improving results. Based on these findings, we provide actionable recommendations for developers and practitioners, emphasizing that successful LLM deployment depends on careful model selection, effective prompt design, and context-aware usage to ensure reliable code generation in real-world software development tasks.
\end{abstract}
]



% \printAffiliationsAndNotice{\mlsysEqualContribution} % otherwise use the standard text.

\section{Introduction}
\footnotetext[1]{Department of Computer Science, Iowa State University, Ames, Iowa, USA. Correspondence to: Le Zhang $<$lezhang@iastate.edu$>$.}
\vspace{0.2in}

Thousands of companies, large and small, are integrating AI tools into their software development workflows. According to recent developer surveys from GitHub and Stack Overflow~\cite{github2024,stackoverflow2024}, over two-thirds of professional developers now use AI-assisted tools in their workflows. While these tools may enhance productivity, growing concerns exist about their impact on code quality and maintainability. GitClear~\cite{gitclear2024} and GeekWire~\cite{geekwire2024} both report sharp increases in “code churn”—the rate at which new code is discarded shortly after creation—suggesting potential declines in maintainability.

These trends raise important questions about the long-term sustainability of LLM-generated code, which may resemble credit card spending: easy to accumulate, but difficult to manage over time~\cite{stanfordcs224}. This highlights the need for a rigorous evaluation of LLM-generated code: how it compares in quality to human-written code, whether significant differences exist between state-of-the-art models, and to what extent these tools can reproduce the design quality of experienced developers.~\cite{boundaries2025, ashraf24}.

As LLMs become deeply embedded in the software development lifecycle, a growing body of research has begun evaluating their capabilities in code generation, debugging, and comprehension~\cite{jiang2024surveylargelanguagemodels, wang2023review, revelo2025, mit2025, medium2025code}. These studies often assess models using criteria such as syntactic correctness, functional accuracy, and security robustness. For example, Chen et al.~\cite{chen2021codex} evaluated Codex across diverse programming tasks, revealing both strengths and limitations. Similarly, Pearce et al.~\cite{pearce2022copilot} identified recurring security vulnerabilities in LLM-generated code, particularly in input validation and sanitization. Dougherty et al.~\cite{dougherty2025proving} pointed out that current LLMs have certain abilities in automated creation of deductively verifiable
code. Other work has explored ways to enhance reliability, such as Nijkamp et al.~\cite{nijkamp2022codegen}, who demonstrated how structured prompting improves robustness across programming languages, and Patel and Yadavally~\cite{patel2025planning}, who examined domain-specific performance in tasks like static analysis and error detection.

While the conversation around LLM-assisted code generation includes both optimism and skepticism~\cite{codeflash2024, ramirez2024, baek2025, liu2023codegeneratedchatgptreally, mohsin2024, fakhoury2024, medium2025python}, there is a clear and urgent need for more rigorous, empirical evaluation to truly understand these tools' capabilities and limitations.

This study presents a comprehensive empirical evaluation of state-of-the-art general-purpose and code-specialized LLMs. We assess their performance on a diverse set of real-world programming problems sourced from LeetCode, a public repository of coding tasks. Our investigation is guided by three primary objectives:

\begin{enumerate}
    \item Evaluate the coding proficiency of modern LLMs across multiple programming languages and problem domains using rigorous metrics.
    \item Quantitatively benchmark their strengths and weaknesses across complementary evaluation dimensions.
    \item Provide practical guidance for integrating LLMs into software development workflows, focusing on model selection, prompt design, and contextual usage.
\end{enumerate}

Through large-scale experiments and quantitative analyses, this paper offers a critical assessment of LLM capabilities and actionable insights for developers and researchers seeking to apply these models in real-world coding environments.

\section{Background}

\vspace{0.1in}
This section provides an overview of recent developments in LLMs and their applications in automated code generation, highlighting key architectures, representative models, and technical advancements relevant to this study.

\vspace{0.1in}
\subsection{Large Language Models (LLMs)}
Recent advances in LLMs have been driven by breakthroughs in transformer architectures, scaling strategies, and training methodologies. These innovations have produced highly capable models across diverse domains. 

Notable open-source examples include Meta's Llama-3.3\cite{dubey2024llama}, designed for multilingual dialogue and coding tasks, and Alibaba's Qwen2.5 series~\cite{Yang2024Qwen25TR,qwen25coder}, specialized in mathematics and programming. DeepSeek-V3\cite{liu2024deepseek} is a general-purpose Mixture-of-Experts (MoE) model, while its variant, DeepSeek-R1~\cite{guo2025deepseek}, also an MoE model, is optimized for complex reasoning and problem-solving tasks.

As for proprietary LLMs, OpenAI's GPT-4.1~\cite{gpt41} offers enhanced reasoning and multilingual capabilities, while Anthropic's Claude 3.7 Sonnet~\cite{claude37} excels in coding and long-context understanding.

These developments reflect a broader shift from general-purpose LLMs toward specialized models, fine-tuned for specific applications such as programming, mathematics, and advanced reasoning.

% \vspace{0.1cm}
\subsection{Code Generation Using LLMs}
LLMs for code generation automate the translation of natural language descriptions into functional source code, a process known as \textbf{natural-language-to-code} (NL2Code)~\cite{jiang2024surveylargelanguagemodels}. This capability has transformed software development by enabling rapid prototyping, automated debugging, and even full application generation~\cite{legit2025risks, sonarsource2024quality}.  

Modern LLMs exhibit strong performance in code generation, with capabilities including: (1) Multi-language support that generates code in Python3, Java, C++, and other languages~\cite{fabric2024llms}; (2) Code translation that converts programs between languages (e.g., Python to JavaScript)~\cite{jiang2024surveylargelanguagemodels}; (3) Debugging and optimization that identifies errors and suggesting improvements~\cite{sonarsource2024quality}.

General-purpose models (e.g., Meta's Llama~\cite{touvron2023llama} and Google's Gemini~\cite{team2023gemini}) demonstrate broad competence but may lack depth in coding-specific tasks~\cite{leetdataset2025, svetkin2024testing}. Code-specialized models (e.g., DeepSeek-Coder~\cite{guo2024deepseek} and Qwen2.5-Coder~\cite{qwen25coder}) are fine-tuned on vast codebases, yielding stronger capabilities for complex coding challenges.

Recent research suggests that LLM-generated code can sometimes surpass human-written solutions in terms of efficiency. However, challenges like hallucinations (i.e., generating incorrect or non-functional code) and security vulnerabilities~\cite{Coignion_2024, jimenez2024evaluation} remnains. To address these issues, advanced techniques such as Retrieval-Augmented Generation (RAG) and Reinforcement Learning from Human Feedback (RLHF) are adopted to improve reliability and robustness.

% \vspace{0.3cm}
\subsection{LeetCode}

LeetCode is a popular online platform hosting thousands of algorithmic and data structure problems, commonly used for technical interview preparation~\cite{leetdataset2025, svetkin2024testing}. 

For evaluating LLMs, LeetCode problems are valuable because of their following characteristics:  
\begin{itemize}  
    \item \textbf{Structured Problem Descriptions}: Each problem provides a clear, well-defined natural language prompt along with formatted input/output examples, ensuring explicit task interpretation~\cite{leetdataset2025}.  
    \item \textbf{Comprehensive Test Cases}: The platform includes extensive test cases, including edge cases and adversarial scenarios, enabling rigorous and objective code evaluation~\cite{leetdataset2025}.  
    \item \textbf{Diverse Difficulty Levels}: Problems are categorized into ``Easy'', ``Medium'', and ``Hard'' tiers, facilitating systematic assessment of LLMs' capabilities across varying skill levels~\cite{svetkin2024testing}.  
\end{itemize}

Previous studies~\cite{doderlein2022piloting, Coignion_2024} are based on older models and they focus on the impact of hyperparameter settings and prompt design. Our study provides a more comprehensive and up-to-date evaluation. We focus on the correctness and overall quality of code generated by the latest state-of-the-art LLMs. By leveraging a large and diverse dataset, broad language coverage, and detailed qualitative analysis, our work delivers a more practical and insightful assessment of modern LLMs' performance in real-world coding scenarios.

\vspace{-0.05in}
\section{EXPERIMENTAL SETUP}
\label{sec:experimental_setup}

To assess the performance of LLMs in generating correct and efficient code, we developed a comprehensive experimental framework. This section outlines the key components of our setup, including models setup, coding question datasets, prompt design, code submission workflow, and evaluation metrics.

\begin{table}[b]
\caption{LLMs used in our study.}
\label{tab1}
\vskip 0.1in
\begin{center}
\begin{small}
\begin{sc}
\resizebox{0.48\textwidth}{!}{
\begin{tabular}{lcccr}
\toprule
Model & \begin{tabular}{@{}c@{}}Size\\ (B)\end{tabular} & \begin{tabular}{@{}c@{}}Release\\ Year\end{tabular} & \begin{tabular}{@{}c@{}}Open\\ Source\end{tabular} \\
\midrule
DeepSeek-R1 & 671 & 2025 & $\surd$ \\
DeepSeek-V3-0324 & 685 & 2025 & $\surd$ \\
Qwen2.5-Coder-32B-Instruct & 32 & 2024 & $\surd$ \\
Llama-3.3-70B-Instruct & 70 & 2024 & $\surd$ \\
GPT-4.1 & 1800 (Est.) & 2025 & $\times$ \\
Claude-3.7-Sonnet & 200 (Est.) & 2025 & $\times$ \\
\bottomrule
\end{tabular}
}
\end{sc}
\end{small}
\end{center}
\vskip -0.2in
\end{table}

\vspace{-0.07in}
\subsection{Models Selection}

In this study, we evaluate six state-of-the-art LLMs to compare their performance in code generation. Table~\ref{tab1} shows the models used in our study and their key specifications. Model size, expressed in billions of parameters, reflects the model’s architectural complexity and potential capability. For simplicity, we refer to them as DeepSeek-R1, DeepSeek-V3, Qwen2.5-Coder, Llama-3.3, GPT-4.1, and Claude-3.7, respectively, throughout the paper.

\vspace{-0.07in}
\subsection{Coding Question Datasets}
\label{subsec:dataset}

\vspace{-0.05in}
We created two datasets to evaluate the capabilities of LLMs in solving programming problems with varying complexity and constraints:

\vspace{-0.1in}

\begin{itemize}
    \item \textbf{Large Dataset (944 Questions)} \\
    This dataset consists of 944 manually picked LeetCode programming questions focusing on data structures and algorithms. The selected problems cover four primary topics: greedy algorithms, sorting, binary search, and tree-based problems. 
    % These are relatively recent additions to the LeetCode question pool and are therefore less likely to have been included in the training data of current LLMs, reducing the risk of data contamination.

    \item \textbf{Small Dataset (202 Questions)} \\
    % This dataset is a subset of the large dataset, includes 202 questions specifically selected for their sensitivity to algorithmic complexity. These problems demand highly optimized solutions, as suboptimal algorithms will fail due to time constraints. By focusing on such challenges, the dataset provides a rigorous benchmark for evaluating the time complexity of LLM-generated code.
    This subset comprises 202 questions selected from the larger dataset based on their sensitivity to algorithmic complexity. These problems require highly optimized solutions, as suboptimal algorithms often fail to meet time constraints. Focusing on such complexity-critical tasks provides a rigorous benchmark for evaluating the efficiency and time complexity of LLM-generated code.

\end{itemize}

LeetCode problems are classified into three difficulty levels: Easy, Medium, and Hard. Table~\ref{tab2} presents the distribution of these difficulty levels across the two datasets used in our study.

LeetCode provides a set of test cases for each coding problem to evaluate the correctness and performance of submitted solutions. The number of test cases per problem varies widely, ranging from a minimum of 7 to a maximum of 9,558. Table~\ref{tab3} presents the distribution of problems based on the number of test cases they include, highlighting the diversity and rigor of the evaluation criteria across different problems.

While our datasets are derived from publicly available sources, the possibility that some problems or their corresponding solutions may have appeared in the pretraining corpora of the evaluated LLMs. To mitigate this risk, we intentionally selected relatively recent problems that are less likely to be included in existing training corpora, avoided using any LeetCode discussion content or official solutions, and treated all prompts as unseen tasks. Although complete elimination of contamination cannot be guaranteed, our methodology was designed to mitigate it to the greatest practical extent.

\begin{table}[b!]
\caption{Problem distribution by difficulty levels.}
\label{tab2}
\vskip 0.1in
\begin{center}
\begin{small}
\begin{sc}
\begin{tabular}{lcccc}
\toprule
Dataset & Easy & Medium & Hard & Total \\
\midrule
Large Dataset & 158 & 555 & 231 & 944 \\
Small Dataset & 0 & 103 & 99 & 202 \\
\bottomrule
\end{tabular}
\end{sc}
\end{small}
\end{center}
\vskip -0.1in
\end{table}

\begin{table}[b!]
\caption{Problem distribution by number of test cases.}
\label{tab3}
\vskip 0.1in
\begin{center}
\begin{small}
\begin{sc}
\resizebox{0.48\textwidth}{!}{
\begin{tabular}{lccccc}
\toprule
Dataset & 1--49 & 50--99 & 100--999 & 1k+ & Total \\
\midrule
Large Dataset & 173 & 328 & 390 & 53 & 944 \\
Small Dataset & 40 & 63 & 88 & 11 & 202 \\
\bottomrule
\end{tabular}
}
\end{sc}
\end{small}
\end{center}
\vskip -0.2in
\end{table}

\vspace{-0.07in}
\subsection{Parameters Configuration}

To ensure high-quality and consistent outputs, we carefully configured the key decoding parameters across all models. Prior studies indicate that setting the top-p parameter between 0.9 and 1.0 generally yields better performance for LLMs~\cite{holtzman2019curious, zheng2023gpt, JanSiml}. In particular, research on LLM-generated LeetCode solutions and DeepSeek models reports that a top-p value of 0.95 enhances coding performance, especially on the Pass@1 metric~\cite{Coignion_2024, guo2025deepseek}. Following these findings, we adopted top-p = 0.95 for all six models to allow a broader sampling distribution and improve solution quality.

Previous work also suggests that higher temperature values can increase success rates in multi-attempt code generation experiments (e.g., Pass@k), but often reduce accuracy in one-shot settings such as Pass@1~\cite{zheng2023gpt}. Elevated temperature introduces greater randomness in token selection, leading to inconsistent or incorrect outputs. Therefore, to maintain stability and reproducibility in our one-shot experiments, we fixed the temperature at 0.1 for all models.

\vspace{-0.07in}
\subsection{Prompt Configuration}
\label{subsec:prompt_setup}

A structured prompt was designed to ensure consistency and clarity throughout the LLMs' code generation process. An example of this prompt format is shown in Figure~\ref{fig:prompt}. The prompt comprises the following components:
\vspace{-0.05in}
\subsubsection{Role Specification}

Each prompt begins by defining the model's role to establish task context. In this case, the model is instructed to act as a software developer and implement a solution in a specified language (e.g., Python). This helps ensure that the output remains focused on executable code rather than general explanations.

\vspace{-0.05in}
\subsubsection{Problem Description and Example}

Next, the Problem Description presents the coding task in natural language, followed by Examples that illustrate the expected input–output behavior. These examples clarify the problem requirements and guide the model toward functionally correct solutions.

\vspace{-0.05in}
\subsubsection{Constraints and Code Snippet}

The Constraints section outlines the valid input ranges and conditions, while the Code Snippet provides a partial template (e.g., a class or function skeleton). This structure helps the model concentrate on implementing the solution logic without redefining unnecessary components.

\vspace{-0.05in}
\subsubsection{Test Cases}

Representative testcases are included to indicate how the solution will be evaluated. These examples reinforce the problem requirements and help the model produce outputs consistent with test expectations.

\vspace{-0.05in}
\subsubsection{Additional Instructions}

Finally, the Additional Instructions specify formatting, style, and content constraints. The model is asked to write executable, well-formatted code that adheres to best practices while avoiding unnecessary commentary.

\begin{figure}[!t]
\centering
\fbox{
    \begin{minipage}{0.45\textwidth}
    \textcolor{blue}{\# Start of Prompt}\\
    
    You are a \textbf{software developer}. Implement a solution in \textbf{Python} for the following coding problem.\\
    
    \textbf{Problem Description:}\\
    \textcolor{black}{Given an integer $num$, repeatedly add all its digits until the result has only one digit, and return it.}\\
    
    \textbf{Example:}\\
    Input: $num = 38$\\
    Output: 2\\
    Explanation: The process is\\
    $38 \rightarrow 3 + 8 \rightarrow 11$, then $11 \rightarrow 1 + 1 \rightarrow 2$\\
    Since 2 has only one digit, return it.\\
    
    \textbf{Constraints:}\\
    $0 \leq num \leq 2^{31} - 1$\\

    \textbf{Code Snippet:}\\
    class Solution:
    
    \qquad def addDigits(self, num: int) $\rightarrow$ int:\\

    \textbf{Testcases:}\\
    $num=38$\\ 
    $num=0$\\

    \textbf{Additional Instructions:}\\
    Follow the input constraints and write your code starting from the given code snippet. Ensure the code is well-formatted and adheres to best practices. Write the executable code only, avoid unnecessary explanations or comments.\\
    
    \textcolor{blue}{\# End of Prompt}
    \end{minipage}
}
\vspace{0.1in}
\caption{Prompt structure example}
\label{fig:prompt}
\end{figure}

\subsection{Code Submission}
\label{subsec:code_submission}

As specified in our prompt design, the LLMs were instructed to produce source code only for each coding problem. These generated solutions were submitted to LeetCode via a custom LeetCode API developed for this study. For every coding problem, we made one submission per model and per programming language. Each submission was evaluated using LeetCode's comprehensive set of test cases, and the resulting submission reports were collected in JSON format for further analysis. To comply with ethical guidelines and minimize any potential disruption to LeetCode's services, we strictly followed the platform's usage policies and terms of conduct.

\subsection{Evaluation Metrics}
\label{subsec:evaluation_metrics}

To assess the performance of LLM-generated code, we employed the following evaluation metrics:

\subsubsection{Pass@1 Metric}
\label{subsubsec:pass@1}

The Pass@1 metric measures the percentage of problems for which a model produces a correct and accepted solution on its first attempt. A solution is considered successful if it compiles, runs without errors, and passes all test cases within LeetCode's time constraints. This metric represents the overall accuracy and practical usability of an LLM in single-shot code generation settings.

Formally, for a dataset containing $N$ programming problems, if $S_i$ indicates the model's first attempt on problem $i$ is accepted ($S_i = 1$) or rejected ($S_i = 0$), the Pass@1 score is computed as:
\[
\text{Pass@1} = \frac{1}{N} \sum_{i=1}^{N} S_i
\]

where $\text{Pass@1} \in [0, 1]$ represents the proportion of problems successfully solved on the first attempt.

\subsubsection{Compile-time Error (CE) Metric} 
\label{subsubsec:CE}

The CE metric reflects the percentage of submissions that fail to compile due to syntax errors such as missing brackets or undeclared variables. A high CE metric indicates weak syntactic correctness.

\subsubsection{Runtime Error (RE) Metric} 
\label{subsubsec:RE}

The RE metric measures the percentage of submissions that compile successfully but fail during execution due to issues such as division by zero or null pointer exceptions. A high RE metric exposes weaknesses in runtime robustness and error handling.

\subsubsection{Functional Failure (FF) Metric} 
\label{subsubsec:FF}

Even when a submission compiles and executes without crashing, it may still produce incorrect outputs. The FF metric quantifies the percentage of submissions that fail at least one test case. High FF values indicate logical flaws or incorrect algorithmic reasoning.

\subsubsection{Algorithmic Suboptimality (AS) Metric} 
\label{subsubsec:AS}

The AS metric measures the percentage of submissions that exceed execution time limits due to inefficient algorithms (i.e., Time-Limit-Exceeded errors on LeetCode). High AS values highlight a model's limitations in generating computationally efficient or optimized solutions.

\subsection{Replication package}
All the artifacts of this study, including our datasets, code, and evaluation results, are available in a public repository anonymously: \\\href{https://figshare.com/s/26448e92798aab34e407}{https://figshare.com/s/26448e92798aab34e407}

\section{Experimental Results} 
\label{sec:results}

To evaluate the effectiveness and robustness of LLMs in solving programming problems, we conducted a comprehensive empirical analysis across multiple dimensions of code quality and performance. Specifically, we assessed six LLMs across five popular programming languages using the large dataset introduced in Section~\ref{subsec:dataset}, measuring key metrics such as Pass@1, compile-time errors, runtime errors, functional failures, and algorithmic suboptimalities as defined in Section~\ref{subsec:evaluation_metrics}. 

We further used the smaller, performance-sensitive dataset of 202 LeetCode problems to examine how prompt engineering, particularly the inclusion of optimization cues, affects each model's ability to generate algorithmically efficient code. Together, these experiments provide a holistic view of model strengths and weaknesses in producing syntactically correct, functionally accurate, and performance-optimized solutions.

\subsection{Pass@1 Metric}

\begin{figure}[!t]
\centerline{\includegraphics[width=.49\textwidth]{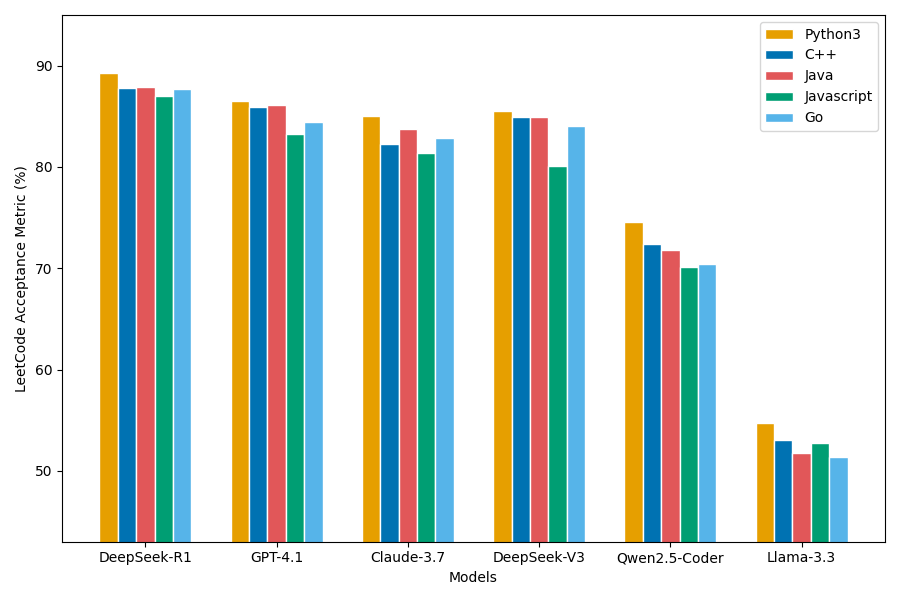}}
\vspace{-0.1in}
\caption{Pass@1 metric by model, higher is better.}
\label{fig5}
\vspace{-0.15in}
\end{figure}

As defined in Section~\ref{subsubsec:pass@1}, Pass@1 represents the proportion of problems solved correctly on the first attempt. Figure~\ref{fig5} presents Pass@1 results by model and language.

DeepSeek-R1 and GPT-4.1 consistently achieve the highest Pass@1 scores across all programming languages, demonstrating superior accuracy, efficiency, and robustness. DeepSeek-V3 and Claude-3.7 also perform well, though slightly behind the top two, indicating solid but somewhat more variable performance.

Qwen2.5-Coder ranks in the middle tier—able to produce correct solutions in many cases, yet less consistent across languages and problem types. In contrast, Llama-3.3 shows the lowest Pass@1 results, suggesting persistent issues with correctness and algorithmic efficiency.

Overall, the findings reveal substantial variation in LLM performance, with only a few models capable of generating high-quality, first-attempt solutions consistently across multiple programming languages.

\subsection{Compile-time Error (CE) Metric}

\begin{figure}[!t]
\centerline{\includegraphics[width=.49\textwidth]{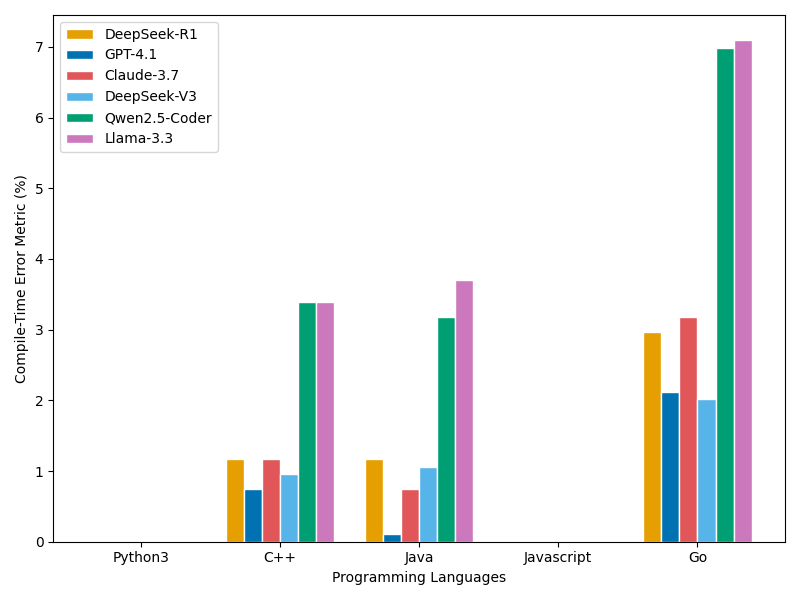}}
\vspace{-0.15in}
\caption{Compile-time Error metric by programming language, lower is better.}
\label{fig1}
\vspace{-0.15in}
\end{figure}

As described in Section~\ref{subsubsec:CE}, the CE metric captures syntactic correctness. Figure~\ref{fig1} shows CE results by language and model.

Across all programming languages, Python3 and JavaScript stand out as the most error-free, with none of the models producing compile-time errors in either language. This likely reflects the simpler, more forgiving syntax of scripting languages and the abundance of related training data, which helps models generate syntactically valid code more consistently.

In contrast, Go tends to exhibit higher CE metric. This may be attributed to several factors, including its relatively recent introduction, having been developed only about 15 years ago, and its stricter syntactic rules compared to other programming languages. Additionally, the limited availability of high-quality Go training data may hinder LLMs from achieving the same level of proficiency as they do with more established languages. 

Models such as Qwen2.5-Coder and Llama-3.3, although strong performers overall, are more prone to compile-time errors in Go. This indicates that strict adherence to language-specific syntactic constraints remains a challenge for current LLMs, particularly in less commonly used or more rigid syntactic languages. Conversely, models like DeepSeek-V3, GPT-4.1, and Claude-3.7 generally produce fewer compile-time errors, indicating stronger syntactic robustness across most languages.

In summary, while some LLMs demonstrate strong performance in executing correct logic, they may still encounter difficulties with syntactic correctness in statically typed or structurally strict languages such as Go, C++, and Java. Python3 and JavaScript, by contrast, appear to be the most reliably supported and least error-prone across all models.

\subsection{Runtime Error (RE) Metric}

\begin{figure}[!t]
\centerline{\includegraphics[width=.49\textwidth]{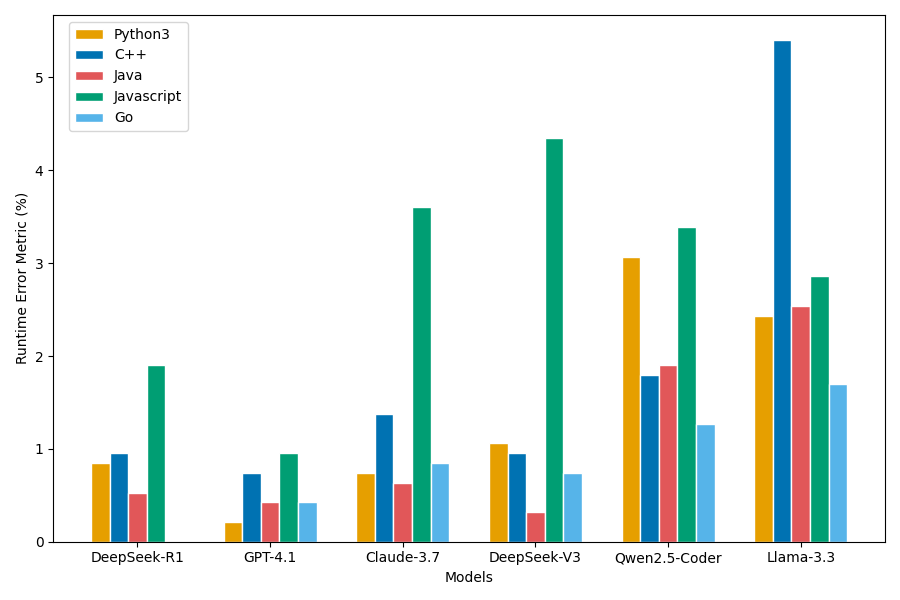}}
\vspace{-0.15in}
\caption{Runtime Error metric by model, lower is better.}
\label{fig2}
\vspace{-0.15in}
\end{figure}

As defined in Section~\ref{subsubsec:RE}, the RE metric reflects runtime robustness and error handling. Figure~\ref{fig2} presents RE results by model and language.

Runtime errors are generally low across most models and programming languages, though distinct patterns emerge. GPT-4.1 exhibits very low RE metrics, which makes it one of the most robust models in terms of execution stability. DeepSeek-R1 also performs well, particularly in Go, where it avoids runtime errors entirely.

Llama-3.3 presents a more mixed performance. It performs well in some languages like Go but shows higher RE metrics in others, especially in C++. This may reflect the challenges of handling languages with stricter type systems and more complex memory management. Similarly, Claude-3.7 and DeepSeek-V3 generally perform reliably but encounter notable runtime issues in JavaScript, suggesting potential weaknesses in managing dynamic or loosely typed languages.

Qwen2.5-Coder exhibits moderate RE metrics that vary across programming languages. While its overall runtime robustness is reasonable, occasional errors, particularly in JavaScript and Python3, highlight areas where improvements are needed for better handling of dynamic execution contexts.

In summary, although all models are somewhat susceptible to runtime errors, those with more extensive general-purpose training tend to exhibit greater robustness. Languages like JavaScript and Python3 present more frequent challenges due to their flexible runtime behavior, whereas statically typed languages such as Java and C++ provide clearer structure but still reveal gaps in model understanding.

\subsection{Functional Failure (FF) Metric}

\begin{figure}[!t]
\centerline{\includegraphics[width=.49\textwidth]{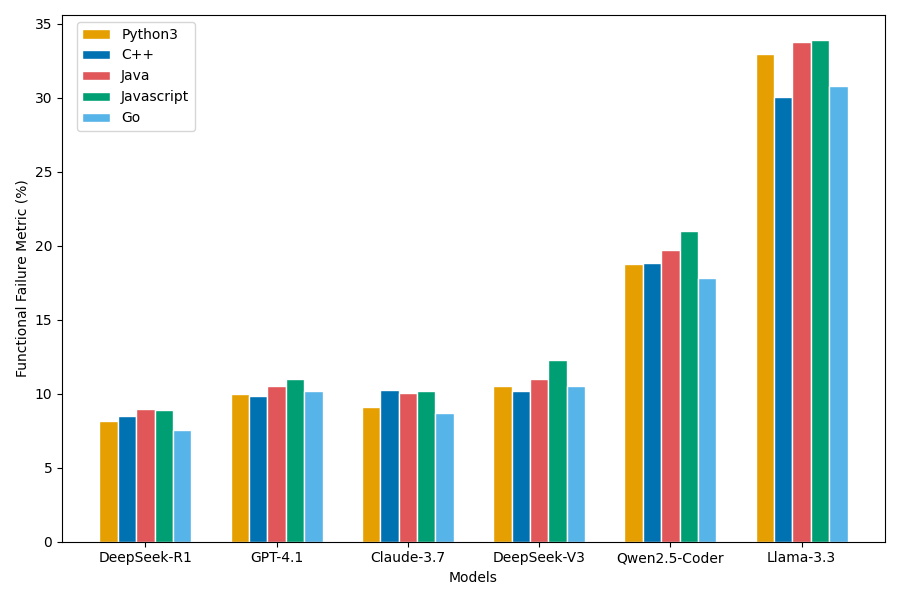}}
\vspace{-0.15in}
\caption{Functional Failure metric by model, lower is better.}
\label{fig3}
\vspace{-0.15in}
\end{figure}

As defined in Section~\ref{subsubsec:FF}, the FF metric captures logical correctness in output. Figure~\ref{fig3} summarizes FF results across models and languages.

DeepSeek-R1 demonstrates the strongest overall performance, with the lowest FF metric across all programming languages, particularly excelling in Python3 and Go. Following closely are Claude-3.7, GPT-4.1, and DeepSeek-V3, which also maintain strong and consistent results across languages, although they are slightly less effective than DeepSeek-R1 in certain cases.

Qwen2.5-Coder falls in the mid-range, showing moderate FF metrics that indicate solid, though not exceptional, code generation capabilities. Its performance is relatively consistent across languages, with only slight variations.

Llama-3.3, while known for its broader general-purpose capabilities, lags behind in this specific coding benchmark. It exhibits noticeably higher FF metrics, particularly in languages like Java and JavaScript, suggesting that it may be less optimized for reliable code generation compared to larger models (e.g., DeepSeek-V3, GPT-4.1) or those more tailored to programming tasks (e.g., Qwen2.5-Coder).

Overall, the results suggest that larger or more specialized models tend to perform better in programming-related evaluations. While there is some variability in performance across languages, the differences between models are more significant than the differences between programming languages.

\subsection{Algorithmic Suboptimality (AS) Metric}

\begin{figure}[!t]
\centerline{\includegraphics[width=.49\textwidth]{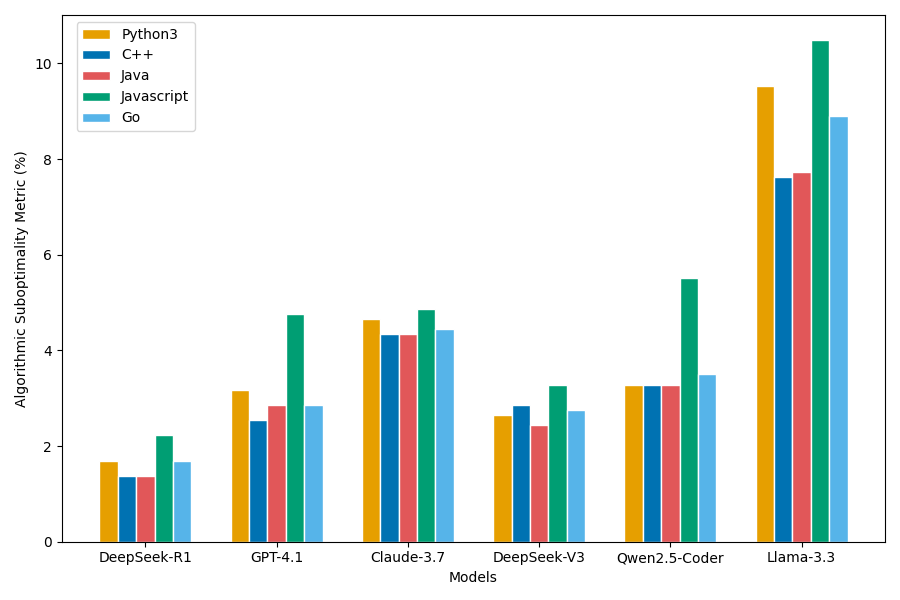}}
\vspace{-0.15in}
\caption{Algorithmic Suboptimality metric by model, lower is better.}
\label{fig4}
\vspace{-0.15in}
\end{figure}

As defined in Section~\ref{subsubsec:AS}, the AS metric quantifies time-limit-exceeded cases indicating inefficiency. Figure~\ref{fig4} compares AS results across models and languages.

Llama-3.3 stands out, but for the wrong reasons. It consistently exhibits the highest AS metric across all programming languages, indicating that while it may generate syntactically and semantically correct code, the underlying algorithms often fail to meet time constraints due to poor optimization.

In contrast, DeepSeek-R1 displays the best performance, with the lowest AS metric across all languages. Its consistent ability to produce code that completes within time limits suggests a stronger grasp of algorithmic optimization, especially when compared to its non-reasoning counterpart, DeepSeek-V3. This highlights the advantages of reasoning-oriented models in efficiency-sensitive coding tasks.

Models such as GPT-4.1 and Qwen2.5-Coder fall in the middle range. Their AS metrics vary across languages, with generally better performance in structured, statically typed languages like C++ and Java, and more challenges in dynamically typed languages like JavaScript and Python3, where the flexibility of the syntax can lead to less efficient implementations. Claude-3.7, on the other hand, demonstrates consistent performance across all languages, indicating a strong level of robustness in cross-language code generation.

Overall, the data underscores that generating functionally correct code is only part of the challenge for LLMs, producing code that is also performant is equally critical. While some models prioritize correctness and robustness, fewer are capable of consistently delivering optimized solutions that meet the performance demands of real-world, large-scale applications.

\subsection{The Impact of Prompt Wording}

\begin{figure}[!t]
\centerline{\includegraphics[width=.49\textwidth]{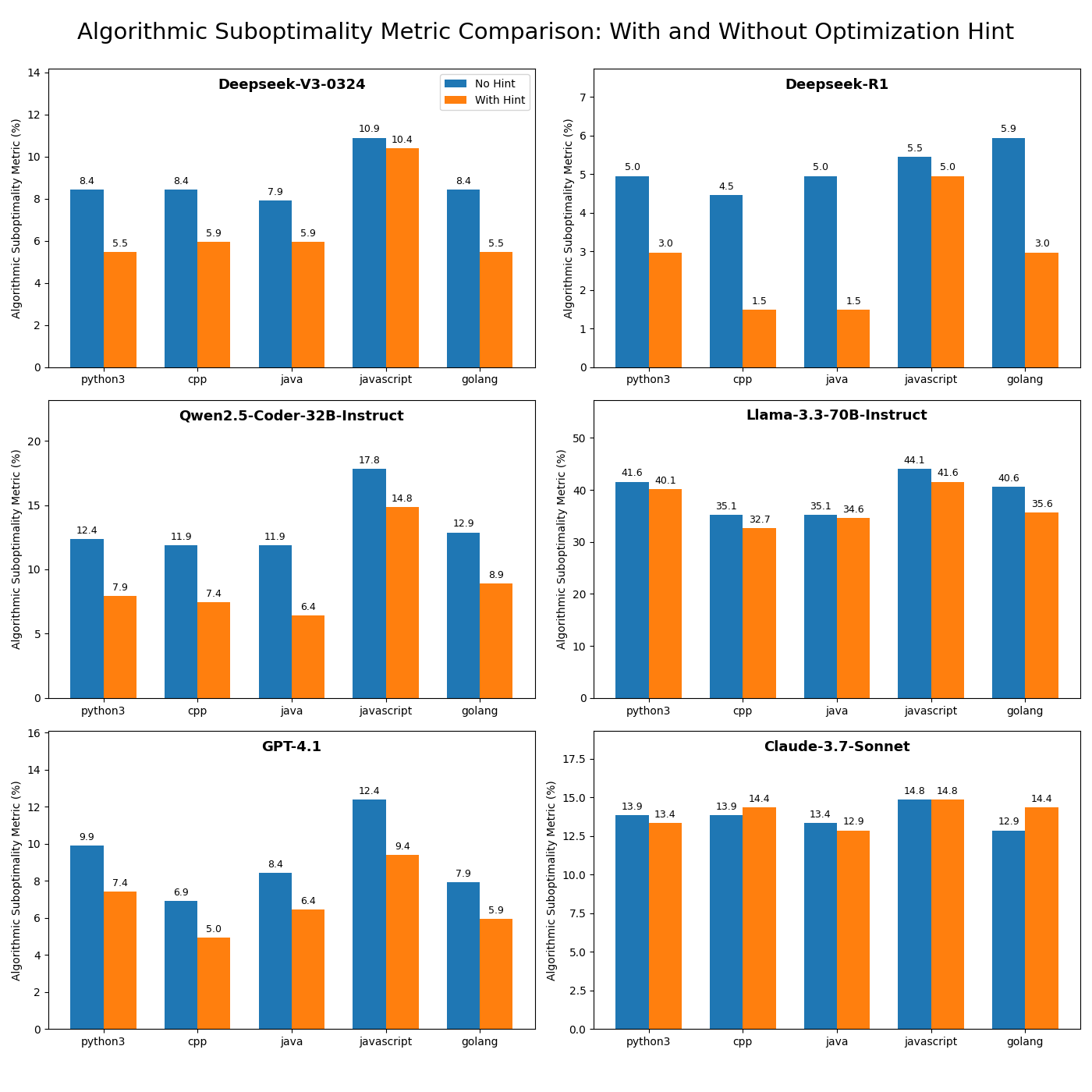}}
\vspace{-0.15in}
\caption{Impact of prompt wording comparison.}
\label{fig6}
\vspace{-0.15in}
\end{figure}

Figure~\ref{fig6} compares the AS metric across six LLMs and five programming languages on the small dataset of 202 LeetCode problems, evaluated with and without an additional optimization prompt:

\textit{“Optimize the time complexity of your algorithm.”}

These problems were specifically selected for their sensitivity to algorithmic efficiency, where optimizing time complexity has a substantial impact on performance outcomes. This experiment examines how models respond to explicit optimization guidance, revealing their sensitivity and adaptability to prompt-based cues aimed at improving algorithmic efficiency. Overall, the inclusion of the optimization hint leads to a reduction in AS values for most models, indicating enhanced solution quality. Notably, DeepSeek-R1 and GPT-4.1 show consistent and substantial improvements across all languages, with DeepSeek-R1 exhibiting the strongest responsiveness to optimization-oriented prompting.

Qwen2.5-Coder and DeepSeek-V3 also benefit from the optimization hint, exhibiting moderate but consistent reductions in AS metrics across all languages. This indicates that these models can leverage explicit guidance to generate more efficient code, even if the improvements are less pronounced than those observed in the top-performing models.

In contrast, Llama-3.3 starts with relatively high AS values and demonstrates only modest improvement when the hint is applied, suggesting that the model struggles to produce time-efficient solutions and is less responsive to prompt-based optimization.

Claude-3.7, meanwhile, shows minimal change in performance with the inclusion of the hint. Its AS metrics remain largely stable across languages and, in some cases, increase slightly, likely due to natural variability or unintended deviations in model output. This behavior may indicate that the model is already performing near its intrinsic optimum or that the prompt exerts limited influence on its behavior.

Overall, the optimization prompt effectively reduces algorithmic suboptimality for several models, particularly DeepSeek-R1, DeepSeek-V3, and GPT-4.1. However, the degree of responsiveness varies, underscoring differences in how each model interprets and adapts to user-provided optimization instructions.

\subsection{Token Usage and Cost}

To complement performance evaluation, we also examine the computational efficiency of each model. This includes analyzing token usage and total operational cost, which together provide insight into the trade-offs between reasoning depth, verbosity, and cost-effectiveness.

\begin{figure}[!t]
\centerline{\includegraphics[width=.49\textwidth]{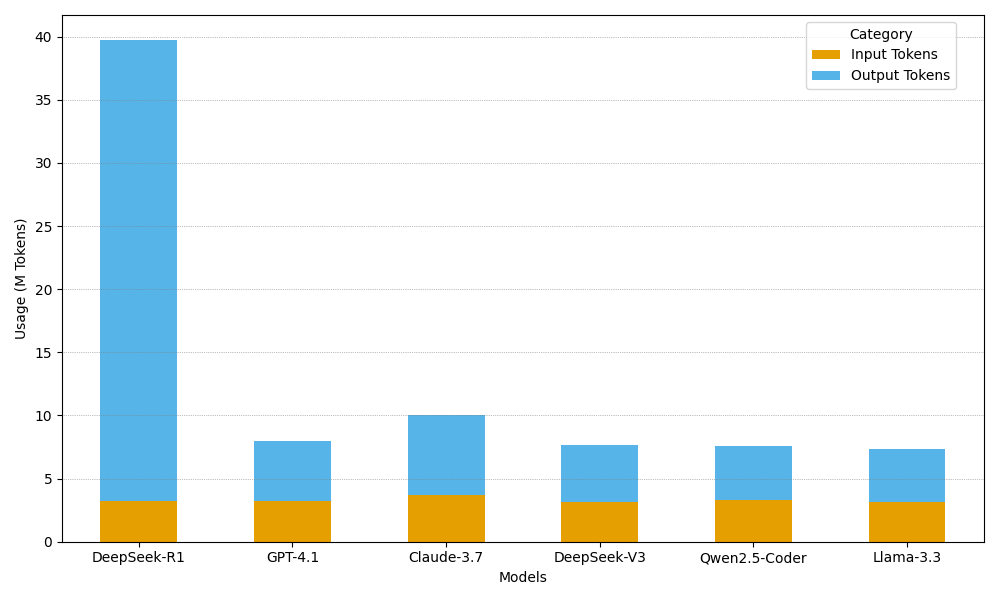}}
\vspace{-0.15in}
\caption{Total token usage across models.}
\label{fig:token_distribution}
\vspace{-0.15in}
\end{figure}

\subsubsection{Token Usage}
Token usage reflects the computational cost of model operation. Data were collected from generating solutions to 944 problems across five programming languages. As shown in Figure~\ref{fig:token_distribution}, input token counts are nearly identical across models due to the standardized prompt format (see Section~\ref{subsec:prompt_setup}), while output token usage varies notably. The reasoning model DeepSeek-R1 consumes substantially more tokens, reflecting its design to allocate extra tokens for intermediate reasoning and more deliberative responses.

Among non-reasoning models, Claude-3.7 exhibits slightly higher token usage, suggesting more verbose outputs. The remaining models, GPT-4.1, DeepSeek-V3, Qwen2.5-Coder, and Llama-3.3, show comparable token efficiency, indicating optimization for concise responses and cost-effective performance.

\begin{figure}[!t]
\centerline{\includegraphics[width=.49\textwidth]{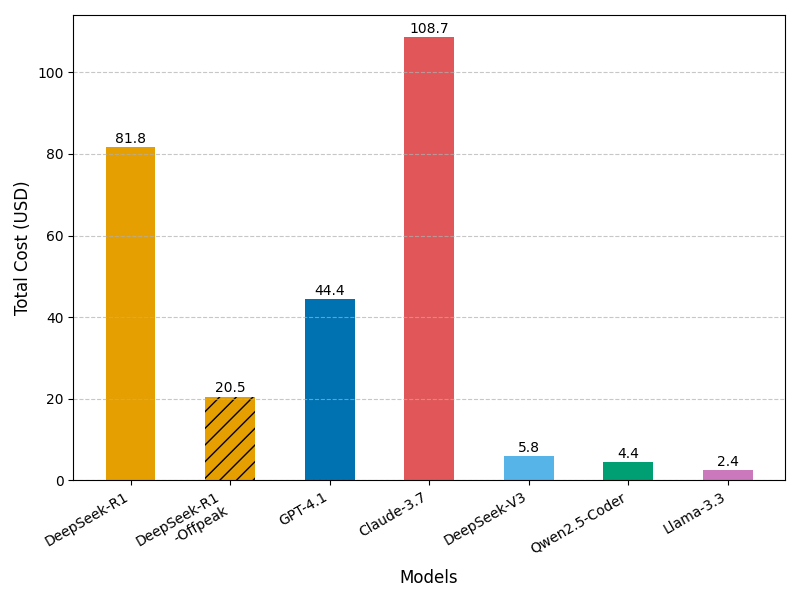}}
\vspace{-0.15in}
\caption{Total cost across models.}
\label{fig:total_cost}
\vspace{-0.15in}
\end{figure}

\subsubsection{Total Cost Analysis}
Figure~\ref{fig:total_cost} compares total operational costs\footnote{API pricing reflects market conditions as of August 2025.} for solving 944 programming problems across five languages. Smaller models such as Qwen2.5-Coder achieve strong cost efficiency while maintaining competitive performance, making them appealing for budget-sensitive deployments.

Despite higher token consumption from Chain of Thought (CoT) reasoning, DeepSeek-R1 maintains moderate overall costs, showing that pricing structure, not token count alone, influences total expense. DeepSeek’s off-peak discount\footnote{DeepSeek's off-peak promotion ended on September 5, 2025.}, in particular, significantly reduced its total cost, illustrating how strategic pricing policies and deployment timing can substantially improve cost efficiency in large-scale applications.

\section{Conclusions and Recommendations}

This study presents a comprehensive evaluation of state-of-the-art LLMs in generating correct and efficient software code, evaluated on a benchmark of LeetCode problems. The results reveal substantial performance differences among models: DeepSeek-R1 and GPT-4.1 consistently outperform others in correctness, efficiency, and robustness, while models such as Llama-3.3 frequently exhibit runtime errors, logical flaws, and algorithmic inefficiencies.

\subsection{Key Findings}
\begin{itemize}
    \item \textbf{Correctness and Robustness}:
    Models like DeepSeek-R1 and GPT-4.1 show high values for Pass@1 metric and low values for CE/RE metrics, reflecting strong syntactic and logical correctness. However, even top models occasionally fail on complex problems or edge cases, underscoring the need for human oversight.
    
    \item \textbf{Efficiency Challenges}:
    Algorithmic suboptimality problems are common, especially in models like Llama-3.3, which often default to brute-force solutions. Prompt engineering, such as using optimization hints, may significantly improve performance in efficiency-driven tasks.
    
    \item \textbf{Language-Specific Nuances}:
    Python3 and JavaScript yield fewer compile and runtime errors, while stricter languages like C++ and Go reveal more limitations. Models sometimes overlook language-specific constraints, such as variable scope or integer overflow, requiring careful review.
\end{itemize}

\subsection{Recommendations for Practitioners}
\begin{itemize}
    \item \textbf{Use High-Performance Models}: For critical tasks, prioritize models like DeepSeek-R1 or GPT-4.1, which combine high correctness with efficient runtime performance.
    
    \item \textbf{Apply Explicit Prompts}: Provide clear optimization hints and constraints (e.g., ``Ensure the solution handles edge cases.'') to guide models toward better solutions.
    
    \item \textbf{Validate and Test}: Treat LLM-generated code as a first draft. Always review for logical correctness, edge cases, and efficiency, especially in performance-sensitive applications.

    \item \textbf{Specialize for Languages}: Be mindful of language-specific pitfalls (e.g., integer overflow in C++) and verify generated code against compiler and runtime constraints.
\end{itemize}

Our findings echo broader critiques of AI's overstated transformative potential. As Acemoglu~\cite{acemoglu2025simple} argues, current AI systems are expected to automate only a small fraction of tasks, around 5\%, and contribute modestly to global productivity. This aligns with our results: even state-of-the-art LLMs exhibit persistent weaknesses in correctness, efficiency, and contextual reasoning. These limitations reinforce the importance of pro-human AI deployment, where technology augments rather than replaces human expertise.

In software engineering, this synergy is particularly evident: LLMs can accelerate development, but human oversight remains essential for debugging, optimizing, and interpreting complex requirements. This insight aligns with Frederick Brooks's conclusion in his Allen Newell address~\cite{brooks1996computer}: ``IA $>$ AI'', intelligence amplification can outperform artificial intelligence alone. Together, these findings advocate for a collaborative paradigm in AI-assisted programming, where human judgment continues to anchor quality and innovation.

\section{Future Work}

This study provides a comprehensive evaluation of state-of-the-art LLMs in generating standalone solutions for coding problems. Future research should explore their ability to modify and evolve existing software code. A promising direction involves leveraging version control histories (e.g., Git commit logs) to train and evaluate LLMs on tasks such as bug fixing, feature addition, and refactoring. Learning from patterns of real-world code evolution, models could provide more context-aware assistance in software maintenance, bridging the gap between standalone code generation and practical software development.

In addition to this line of inquiry, future studies should also investigate the interactive coding dynamics between humans and LLMs. Our current evaluation emphasizes the Pass@1 metric, which measures performance on the first attempt. However, in practice, developers often engage in iterative interactions with LLMs, refining code through feedback from compilers, test cases, or runtime outputs. Investigating such multi-turn, feedback-driven workflows represents an important avenue for future work to better reflect real-world coding practices.

% \section*{Acknowledgements}

% \textbf{Do not} include acknowledgements in the initial version of
% the paper submitted for blind review.

% If a paper is accepted, the final camera-ready version can (and
% probably should) include acknowledgements. In this case, please
% place such acknowledgements in an unnumbered section at the
% end of the paper. Typically, this will include thanks to reviewers
% who gave useful comments, to colleagues who contributed to the ideas,
% and to funding agencies and corporate sponsors that provided financial

\nocite{langley00}

\bibliography{paper}
\bibliographystyle{mlsys2025}

%%%%%%%%%%%%%%%%%%%%%%%%%%%%%%%%%%%%%%%%%%%%%%%%%%%%%%%%%%%%%%%%%%%%%%%%%%%%%%%
%%%%%%%%%%%%%%%%%%%%%%%%%%%%%%%%%%%%%%%%%%%%%%%%%%%%%%%%%%%%%%%%%%%%%%%%%%%%%%%
% SUPPLEMENTAL CONTENT AS APPENDIX AFTER REFERENCES
%%%%%%%%%%%%%%%%%%%%%%%%%%%%%%%%%%%%%%%%%%%%%%%%%%%%%%%%%%%%%%%%%%%%%%%%%%%%%%%
%%%%%%%%%%%%%%%%%%%%%%%%%%%%%%%%%%%%%%%%%%%%%%%%%%%%%%%%%%%%%%%%%%%%%%%%%%%%%%%

% \appendix

\begin{onecolumn}
\centering
\appendix
\section{Data Tables For Experimental Results}

\vspace{-0.2in}

\begin{table*}[h!]
\begin{center}
\caption{Model Performance Data Across Programming Languages (Large Dataset, 944 problems).}
\label{tab-performance}
% \vskip 0.15in
\begin{small}
\begin{sc}
\resizebox{\textwidth}{!}{
\begin{tabular}{lccccccc}
\toprule
Metric & Language & DeepSeek-R1 & GPT-4.1 & Claude-3.7 & DeepSeek-V3 & Qwen2.5-Coder & Llama-3.3 \\
\midrule
\multirow{5}{*}{Pass@1} 
& Python & \textbf{89.30} & 86.55 & 85.06 & 85.49 & 74.58 & 54.77 \\
& C++ & \textbf{87.82} & 85.91 & 82.31 & 84.96 & 72.35 & 53.07 \\
& Java & \textbf{87.92} & 86.12 & 83.79 & 84.96 & 71.82 & 51.80 \\
& Javascript & \textbf{86.97} & 83.26 & 81.36 & 80.08 & 70.13 & 52.75 \\
& Go & \textbf{87.71} & 84.43 & 82.84 & 84.00 & 70.44 & 51.38 \\
\midrule
\multirow{5}{*}{\begin{tabular}{@{}c@{}}Compile-time\\ Error (\%)\end{tabular}}
& Python & 0.00 & 0.00 & 0.00 & 0.00 & 0.00 & 0.00 \\
& C++ & 1.17 & \textbf{0.74} & 1.17 & 0.95 & 3.40 & 3.40 \\
& Java & 1.17 & \textbf{0.11} & 0.74 & 1.06 & 3.18 & 3.71 \\
& Javascript & 0.00 & 0.00 & 0.00 & 0.00 & 0.00 & 0.00 \\
& Go & 2.97 & 2.12 & 3.18 & \textbf{2.01} & 6.99 & 7.10 \\
\midrule
\multirow{5}{*}{\begin{tabular}{@{}c@{}}Runtime\\ Error (\%)\end{tabular}}
& Python & 0.85 & \textbf{0.21} & 0.74 & 1.06 & 3.07 & 2.44 \\
& C++ & 0.95 & \textbf{0.74} & 1.38 & 0.95 & 1.80 & 5.40 \\
& Java & 0.53 & 0.42 & 0.64 & \textbf{0.32} & 1.91 & 2.54 \\
& Javascript & 1.91 & \textbf{0.95} & 3.60 & 4.34 & 3.39 & 2.86 \\
& Go & \textbf{0.00} & 0.42 & 0.85 & 0.74 & 1.27 & 1.69 \\
\midrule
\multirow{5}{*}{\begin{tabular}{@{}c@{}}Functional\\ Failure (\%)\end{tabular}}
& Python & \textbf{8.16} & 9.96 & 9.11 & 10.49 & 18.75 & 32.94 \\
& C++ & \textbf{8.47} & 9.85 & 10.28 & 10.17 & 18.86 & 30.08 \\
& Java & \textbf{9.00} & 10.49 & 10.06 & 11.02 & 19.70 & 33.79 \\
& Javascript & \textbf{8.90} & 11.02 & 10.17 & 12.29 & 20.97 & 33.90 \\
& Go & \textbf{7.52} & 10.17 & 8.69 & 10.49 & 17.80 & 30.83 \\
\midrule
\multirow{5}{*}{\begin{tabular}{@{}c@{}}Algorithmic\\ Suboptimality\\ (\%)\end{tabular}}
& Python & \textbf{1.70} & 3.18 & 4.66 & 2.65 & 3.28 & 9.53 \\
& C++ & \textbf{1.38} & 2.54 & 4.34 & 2.86 & 3.28 & 7.63 \\
& Java & \textbf{1.38} & 2.86 & 4.34 & 2.44 & 3.28 & 7.73 \\
& Javascript & \textbf{2.23} & 4.77 & 4.87 & 3.28 & 5.51 & 10.49 \\
& Go & \textbf{1.70} & 2.86 & 4.45 & 2.75 & 3.50 & 8.90 \\
\bottomrule
\end{tabular}
}
\end{sc}
\end{small}
\end{center}
% \vspace{-0.3in}
\end{table*}

\vspace{0.3in}

\begin{table*}[h!]
\begin{center}
\caption{Algorithmic Suboptimality Data With and Without Extra Hints (Small Dataset, 202 problems).}
\label{tab-hints}
% \vskip 0.15in
\begin{small}
\begin{sc}
\resizebox{\textwidth}{!}{
\begin{tabular}{lcccccccccccc}
\toprule
% Language & \multicolumn{2}{c}{DeepSeek-R1} & \multicolumn{2}{c}{GPT-4.1} & \multicolumn{2}{c}{Claude-3.7} & \multicolumn{2}{c}{DeepSeek-V3} & \multicolumn{2}{c}{Qwen2.5-Coder} & \multicolumn{2}{c}{Llama-3.3} \\
% \cmidrule(lr){2-3} \cmidrule(lr){4-5} \cmidrule(lr){6-7} \cmidrule(lr){8-9} \cmidrule(lr){10-11} \cmidrule(lr){12-13}
% Extra Hint & $\times$ & $\surd$ & $\times$ & $\surd$ & $\times$ & $\surd$ & $\times$ & $\surd$ & $\times$ & $\surd$ & $\times$ & $\surd$ \\

\multirow{2}{*}{Language} & \multicolumn{2}{c}{DeepSeek-R1} & \multicolumn{2}{c}{GPT-4.1} & \multicolumn{2}{c}{Claude-3.7} & \multicolumn{2}{c}{DeepSeek-V3} & \multicolumn{2}{c}{Qwen2.5-Coder} & \multicolumn{2}{c}{Llama-3.3} \\
\cmidrule(lr){2-3} \cmidrule(lr){4-5} \cmidrule(lr){6-7} \cmidrule(lr){8-9} \cmidrule(lr){10-11} \cmidrule(lr){12-13}
 & Hint$\times$ & Hint$\surd$ & Hint$\times$ & Hint$\surd$ & Hint$\times$ & Hint$\surd$ & Hint$\times$ & Hint$\surd$ & Hint$\times$ & Hint$\surd$ & Hint$\times$ & Hint$\surd$ \\
\midrule
Python & 4.95 & 2.97 & 9.90 & 7.43 & 13.86 & 13.37 & 8.42 & 5.45 & 12.38 & 7.92 & 41.58 & 40.10 \\
C++ & 4.46 & 1.49 & 6.93 & 4.95 & 13.86 & 14.36 & 8.42 & 5.94 & 11.88 & 7.43 & 35.15 & 32.67 \\
Java & 4.95 & 1.49 & 8.42 & 6.44 & 13.37 & 12.87 & 7.92 & 5.94 & 11.88 & 6.44 & 35.15 & 34.65 \\
Javascript & 5.45 & 4.95 & 12.38 & 9.41 & 14.85 & 14.85 & 10.89 & 10.40 & 17.82 & 14.85 & 44.06 & 41.58 \\
Go & 5.94 & 2.97 & 7.92 & 5.94 & 12.87 & 14.36 & 8.42 & 5.45 & 12.87 & 8.91 & 40.59 & 35.64 \\
\bottomrule
\end{tabular}
}
\end{sc}
\end{small}
\end{center}
% \vspace{-0.2in}
\end{table*}

% \vspace{-0.2in}

\begin{table}[h!]
\caption{Total Token Usage and Cost (Generating Solutions for 944 Problems in 5 Programming Languages).}
\label{tab-cost-analysis}
% \vskip 0.15in
\begin{center}
\begin{small}
\begin{sc}
\resizebox{\textwidth}{!}{
\begin{tabular}{clcccccccccccc}
\toprule
& & \multicolumn{2}{c}{DeepSeek-R1} & \multicolumn{2}{c}{GPT-4.1} & \multicolumn{2}{c}{Claude-3.7} & \multicolumn{2}{c}{DeepSeek-V3} & \multicolumn{2}{c}{Qwen2.5-Coder} & \multicolumn{2}{c}{Llama-3.3} \\
\cmidrule(lr){3-4} \cmidrule(lr){5-6} \cmidrule(lr){7-8} \cmidrule(lr){9-10} \cmidrule(lr){11-12} \cmidrule(lr){13-14}
& & Input & Output & Input & Output & Input & Output & Input & Output & Input & Output & Input & Output \\
\midrule
\multicolumn{2}{l}{Usage (M tokens)} & 3.21 & 36.52 & 3.23 & 4.74 & 3.72 & 6.31 & 3.17 & 4.53 & 3.27 & 4.32 & 3.17 & 4.14 \\
\midrule
\multirow{2}{*}{\begin{tabular}{@{}c@{}}API Pricing\footnotemark[1]\\ (USD/M token)\end{tabular}} & Regular & 0.55 & 2.19 & 2 & 8 & 3.75 & 15 & 0.27 & 1.1 & 0.27 & 0.82 & 0.23 & 0.4 \\
% & Off-Peak\footnotemark[2] & 0.135 & 0.55 & $\times$ & $\times$ & $\times$ & $\times$ & 0.135 & 0.55 & $\times$ & $\times$ & $\times$ & $\times$ \\
& Off-Peak\footnotemark[2] & 0.135 & 0.55 & - & - & - & - & 0.135 & 0.55 & - & - & - & - \\
\midrule
\multirow{2}{*}{Cost (USD)} & Regular & 1.77 & 79.98 & 6.45 & 37.95 & 13.95 & 94.69 & 0.86 & 4.98 & 0.88 & 3.54 & 0.73 & 1.66 \\
% & Off-Peak & 0.43 & 20.09 & $\times$ & $\times$ & $\times$ & $\times$ & 0.43 & 0.61 & $\times$ & $\times$ & $\times$ & $\times$ \\
& Off-Peak & 0.43 & 20.09 & - & - & - & - & 0.43 & 0.61 & - & - & - & - \\
\midrule
\multirow{2}{*}{\begin{tabular}{@{}c@{}}Total Cost\\ (USD)\end{tabular}} & Regular & \multicolumn{2}{c}{81.75} & \multicolumn{2}{c}{44.41} & \multicolumn{2}{c}{108.65} & \multicolumn{2}{c}{5.84} & \multicolumn{2}{c}{4.42} & \multicolumn{2}{c}{2.39} \\
% & Off-Peak & \multicolumn{2}{c}{20.52} & \multicolumn{2}{c}{$\times$} & \multicolumn{2}{c}{$\times$} & \multicolumn{2}{c}{1.04} & \multicolumn{2}{c}{$\times$} & \multicolumn{2}{c}{$\times$} \\
& Off-Peak & \multicolumn{2}{c}{20.52} & \multicolumn{2}{c}{-} & \multicolumn{2}{c}{-} & \multicolumn{2}{c}{1.04} & \multicolumn{2}{c}{-} & \multicolumn{2}{c}{-} \\
\bottomrule
\end{tabular}
}
\end{sc}
\end{small}
\end{center}
\vspace{-0.3in}
\end{table}

\footnotetext[1]{API pricing reflects market conditions as of August, 2025, when this study was conducted.}
\footnotetext[2]{DeepSeek's off-peak promotion ended on September 5, 2025.}
\end{onecolumn}

% %
% Put anything that you might normally include after the references as an appendix here, {\it not in a separate supplementary file}. Upload your final camera-ready as a single pdf, including all appendices.

%%%%%%%%%%%%%%%%%%%%%%%%%%%%%%%%%%%%%%%%%%%%%%%%%%%%%%%%%%%%%%%%%%%%%%%%%%%%%%%
%%%%%%%%%%%%%%%%%%%%%%%%%%%%%%%%%%%%%%%%%%%%%%%%%%%%%%%%%%%%%%%%%%%%%%%%%%%%%%%

\end{document}